\documentstyle[11pt]{article}

\newcommand{\prepr}[1] {\begin{flushright} {\bf #1} \end{flushright}
  \vskip 1.5cm}
\newcommand{\titul}[1] {\begin{center}{\large\bf #1 }
\end{center}\vskip 1.cm}

\newcommand{\abstr}[1] {{\begin{center} \vskip .5cm {\bf Abstract
                        \vspace{0pt}} \end{center}}\begin{quote} #1
                        \end{quote}}

\begin{document}

\begin{titlepage}
\prepr{INR-947/97\\JINR E2-97-194\\US-FT/20-97}
\titul{Next-to-next-to-leading order  QCD
analysis  of the revised CCFR data for  $xF_3$
structure function and the higher twist contributions
}
\begin{center}
{\bf A.L. Kataev$^{1}$,~~A.V. Kotikov$^{2}$,~~~G. Parente$^{3}$ and
A.V. Sidorov$^{4}$}\\ [1cm]

$^{1}$ {\em Institute for Nuclear Research of the Academy
of Sciences of Russia,\\ 117312 Moscow, Russia} \\
 $^{2}$ {\em Particle Physics Laboratory, Joint Institute for Nuclear
Research,\\ 141980 Dubna, Russia}\\
$^{3}$ {\em Department of Particle Physics, University of Santiago
de Compostela,\\ 15706 Santiago de Compostela, Spain} \\
$^{4}$ {\em Bogoliubov Laboratory of Theoretical Physics,
Joint Institute for Nuclear Research,\\ 141980 Dubna, Russia}

\end{center}

\abstr{We present the results of the next-to-next-to-leading order
QCD analysis of the recently revised
experimental data of the CCFR collaboration
for the $xF_3$   structure function
using the Jacobi polynomial expansion method.
The effects of the higher twist contributions are included
into the fits following the infrared renormalon motivated
model. The special attention is paid to the checks of the predictive
abilities of the infrared renormalon model and to the independent
extractions of the $x$-dependent shape of the twist-4 contributions
to the $xF_3$ structure function in the process of the leading order,
next-to-leading order and next-to-next-to-leading order fits of the
revised CCFR data. We stress that at the next-to-next-to-leading order
the results for $\alpha_s(M_Z)$ turn out to be almost nonsensitive to the
higher-twist terms. We obtain the following result
$\alpha_s^{NNLO}(M_Z)=0.117\pm 0.002(stat)\pm 0.005(syst)\pm 0.003 (theory)$.
The comparison of the outcomes of our next-to-leading order
and next-to-next-to-leading order analysis
indicate that the theoretical QCD uncertainties were
underestimated in the process of the next-to-leading order
determination of $\alpha_s(M_Z)$, made recently by the CCFR
collaboration itself.}

\hspace{2cm}
{\em Presented in part at the QCD session of Recontres des Moriond-97
(March, 1997).}

\end{titlepage}

{\bf 1.}~The detailed QCD studies of the behavior of the structure
functions of the non-polarized deep-inelastic scattering (DIS) are
still very important both from the phenomenological and theoretical
points of view (see e.g. the reviews \cite{QCD20,Mallot}).
The revision \cite{CCFRrev} of the experimental data of the CCFR
collaboration, previously presented in Ref.\cite{CCFRold} put
on the agenda the reconsideration of the results of the fits of the
old CCFR data, performed at the next-to-next-to-leading order (NNLO)
of perturbative QCD both without higher twist \cite{our4} and with
higher twist contributions \cite{Sasha} using the Jacobi polynomial
expansion method.
The results of application of this method for the extraction
of the $x$-shape of the twist-4 contributions from the NNLO
fits of the combined $xF_3$ data (besides the ones analyzed
in Ref.\cite{Sasha}) can be found in Ref.\cite{Sasha2}.
The Jacobi polynomial approach
was developed in
Refs.\cite{PS}-\cite{Dubna} and successfully used in the process
of the next-to-leading order (NLO) fits \cite{BCDMS}-\cite{ShMS}
and NNLO fits \cite{PKK,our4,Sasha,Sasha2}
of different experimental data.

It should be stressed that at the NLO level the results obtained
turned out to be in good agreement with the outcomes of applications
of the traditional variant of the
DGLAP method \cite{DGLAP}, based on the solution of the
integro-differential equation
(compare the results
of Refs.\cite{KatSid,our4} and Ref.\cite{CCFRold} in the case of
neutrino-nucleon DIS and of Refs.\cite{BCDMS,KPS} with the ones of
Ref.\cite{VM} in the case of the analysis of the data for the $F_2$
structure function, obtained by the BCDMS and SLAC collaborations).
However, the Jacobi polynomial expansion method is also giving the
possibility to include into the QCD fits the existing information
about the NNLO approximations of the coefficient functions and
anomalous dimensions of the finite number of Mellin moments of the
DIS structure functions, calculated analytically in Refs.\cite{VZ1}-
\cite{VZ2} and Ref.\cite{LRV} respectively (for the detailed
methodological description of the application of the Jacobi
polynomial expansion machinery at the NNLO see Ref.\cite{our4}).
Therefore, in view of the current lack of information about the
NNLO corrections to the Alarelli-Parisi kernel, the Jacobi
polynomial expansion method turns out to be the unique method
of including selfconsistently the available NNLO perturbative QCD
effects into the fits of the DIS structure functions data.

In the present work we demonstrate once more the compatibility
of the DGLAP method and the Jacobi polynomial expansion approach
at the NLO level and extract the values of the parameter
$\Lambda_{\overline{MS}}^{(4)}$ from the revised CCFR data
of Ref.\cite{CCFRrev} (hereafter called as the CCFR'97 data)
with taking into account
target mass corrections and twist-4 effects, which were introduced
in our analysis using the concept of the infrared renormalon (IRR)
model of Ref.\cite{DW}. At the next stage we include into our
analysis the NNLO perturbative QCD corrections.
We  extract the $x$-shape
of the twist-4 contributions to the $xF_3$ structure function
at the leading order (LO), NLO and NNLO and compare it with the IRR
model predictions of Ref.\cite{DW}. We also get
the values of $\alpha_s(M_Z)$ both at the NLO and NNLO and
compare them with the ones obtained by the CCFR collaboration in
Ref.\cite{CCFRrev}.

{\bf 2.}~The application of the Jacobi polynomial expansion machinery
at the NLO and NNLO was described in detail in
Refs.\cite{PS}-\cite{Dubna} and Refs.\cite{PKK,our4} correspondingly.
Nevertheless we will introduce again some basic definitions.
The NNLO QCD $Q^2$-evolution of the moments of
the $xF_3$ structure
function $M_n^{NS}(Q^2)=\int_0^1 x^{n-1}F_3(x,Q^2)dx$
(where $n=2,3,4,...$) is determined by the solution
of the  renormalization group equation as
\begin{equation}
\frac{M_n^{NS}(Q^2)}{M_n^{NS}(Q_0^2)}=\bigg[\frac{\alpha_s(Q_0^2)}
{\alpha_s(Q^2)}\bigg]^{\gamma_{NS}^{(0)}(n)/2\beta_0}
H_n(Q_0^2,Q^2)~~~, n\geq 2
\label{1}
\end{equation}
where
\begin{equation}
H_n(Q_0^2,Q^2)=\frac{
[1+p(n)A_s(Q^2)+q(n)(A_s(Q^2))^2]}
{[1+p(n)A_s(Q_0^2)+q(n)(A_s(Q_0^2))^2]}\frac{C_{NS}^{(n)}(A_s(Q^2))}
{C_{NS}^{(n)}(A_s(Q_0^2))}
\label{mq}
\end{equation}
and $C_{NS}^{(n)}(A_s(Q^2))=1+C^{(1)}(n)A_s(Q^2)+C^{(2)}(n)A_s^2(Q^2)$,
$A_s(Q^2)=\alpha_s(Q^2)/(4\pi)$, $p(n)$, $q(n)$ are related
to the coefficients of the anomalous dimension functions
$\gamma_{NS}^{(n)}(A_s)$ and the QCD $\beta$-function
by the equations defined in Ref.\cite{our4} and the numerical
expressions of the coefficients $\gamma_{NS,xF_3}^{(1)}(n)$,
$\gamma_{NS}^{(2)}(n)$, $C_{xF_3}^{(1)}(n)$, $C_{xF_3}^{(2)}(n)$
for $f=4$ are given in our previous paper of Ref.\cite{our4}.

Having at hand the QCD expressions for the Mellin moments
$M_n(Q^2)$ we can reconstruct the structure function $xF_3(x,Q^2)$
using the Jacobi polynomial expansion method:
\begin{equation}
xF_{3}^{N_{max}}(x,Q^2)=x^{\alpha}(1-x)^{\beta}\sum_{n=0}^{N_{max}}
\Theta_n ^{\alpha , \beta}
(x)\sum_{j=0}^{n}c_{j}^{(n)}{(\alpha ,\beta )}
M_{j+2,xF_3} \left ( Q^{2}\right ),
\label{e7}
\end{equation}
where $\Theta_n^{\alpha,\beta}$ are the Jacobi polynomials
and $\alpha,\beta$ are their parameters, fixed by the condition
of the requirement of the minimization of the error of the
reconstruction of the
structure functions (see Ref.\cite{Dubna} for details)
and $M_n(Q^2)$ is the QCD expression for the Mellin moments,
corrected by the target mass contributions and twist-4 effects.
Throughout this paper we will use the following expression
\begin{equation}
M_{n}(Q^2)=M_n^{NS}(Q^2)+\frac{n(n+1)}{n+2}\frac{M_{nucl.}^2}{Q^2}
M_{n+2}^{NS}(Q^2)
+\tilde{C}(n)M_n^{NS}(Q^2)\frac{A_2^{'}}{Q^2}+O(\frac{1}{Q^4})
\label{m3}
\end{equation}
where $A_2^{'}$ is the free parameter, $M_{nucl}$ is the mass of the
nucleon, $\tilde{C}(n)=-n-4+4/n+2/(n+1)+4/(n+2)+4S_1(n)$
$(S_1(n)=\sum_{j=1}^{n-1}1/j)$ is determined by the IRR model
calculations of Ref.\cite{DW} (for the related studies of the
IRR model predictions for the non-singlet structure functions
of DIS see Ref.\cite{AZ}). It is worth to mention here
that the IRR model calculations are known to be complimentary to
the ones, based on
the dispersion approach for the definition
of the running coupling, developed  e.g. in
the papers of Ref.\cite{anal}. Note
also, that the effects of the $O(1/Q^4)$-contributions in Eq.(4)
come from two sources-the target mass effects (see Ref.\cite{our4})
and the twist-6 contributions, which can be extracted from the IRR
model calculations of Ref.\cite{DW}. However, the concrete fits of
the experimental data demonstrate that the $O(1/Q^4)$-effects of the
target mass corrections are negligibly small \cite{our4}. The similar
feature was revealed while taking into account twist-6 contributions
in the process of the fits of the  CCFR'97 data (see the second
citation in Ref.\cite{CCFRrev}). Therefore, in our concrete studies
we will neglect the $O(1/Q^4)$-contributions as the whole.
We will also neglect the problem of defining selfconsistently
the perturbative contribution to the coefficient function of the
twist-4 correction in Eq.(4), which generally speaking differs
from $M_n^{NS}(Q^2)$.

One
more theoretical input of our analysis is the factorization-scale
dependent factor $M_n^{NS}(Q_0^2)$ in Eq.(1). Throughout this work
we will use the $\overline{MS}$-factorization prescription, fixing
$M_n^{NS}(Q_0^2)$ as
\begin{equation}
M_n^{NS}(Q_0^2)=\int_0^1 dx x^{n-2}A(Q_0^2)x^{b(Q_0^2)}(1-x)^{c(Q_0^2)}
(1+\gamma(Q_0^2)x)
\label{fac}
\end{equation}
where $Q_0^2$ is the relevant factorization scale. Note also
that we will take in Eq.(3) $N_{max}=10$ for performing LO and
NLO fits, while our NNLO fits will be done in the case of
$N_{max}=6$. At the NNLO we will use the calculated in Ref.\cite{LRV}
NNLO contributions $\gamma_{NS}^{(2)}(2)$, $\gamma_{NS}^{(2)}(4)$,
$\gamma_{NS}^{(2)}(6)$, $\gamma_{NS}^{(2)}(8)$,
$\gamma_{NS}^{(2)}(10)$ to the anomalous dimension functions and
the estimated by the procedure of smooth
interpolation
results for $\gamma_{NS}^{(2)}(3)$, $\gamma_{NS}^{(2)}(5)$,
$\gamma_{NS}^{(2)}(7)$, $\gamma_{NS}^{(2)}(9)$ \cite{our4}.
It is worth to mention here that the similar procedure  was
previously applied in the process of the NNLO non-singlet fits
of the BCDMS data in Ref.\cite{PKK}.
However, we will avoid the application of the extrapolation of
the expression for $\gamma_{NS}^{(2)}(n)$ at $n>10$.

To extract the $x$-shape of the twist-4 contributions we will
simulate the $xF_3$ structure function by the following expression
\begin{equation}
xF_3(x,Q^2)=xF_3^{LT}(x,Q^2)+\frac{h(x)}{Q^2}
\label{twist}
\end{equation}
where the $Q^2$-dependence of the leading twist term
$xF_3^{LT}(x,Q^2)$ in Eq.(6) is determined by perturbative QCD
(including the proper treatment of the target mass corrections)
and the problem of fixing the $Q^2$-dependence of the
coefficient function of the twist-4 contribution is not considered. The
constants $h(x_i)$ (one per each $x$-bin) are parametrizing the
$x$-dependence of the twist-4 contribution, which in accordance
with the theoretical predictions of Refs.\cite{Br,Nason} in the limit
$x\rightarrow 1$ can be modeled as $h(x)\rightarrow (1-x)^{c-1}$,
where the parameter $c$ is introduced in Eq.(5) and like the
parameters $h(x_i)$, $A,b,\gamma$ and the QCD scale parameter
$\Lambda_{\overline{MS}}^{(4)}$ can be determined by fitting the
concrete sets of the experimental data for
the $xF_3$ structure function.
Note, that the parameter $\Lambda_{\overline{MS}}^{(4)}$ is related
to the expression for the QCD coupling constant $\alpha_s$ via the
standard expansion in inverse powers of $ln(Q^2/
\Lambda_{\overline{MS}}^{(4)~2})$.

{\bf 2.}~At the first stage we performed the fits of the CCFR'97
data of Ref.\cite{CCFRrev}
for the cuts of these data at $x<0.7$ using the value of the
factorization point $Q_0^2=5~GeV^2$
with the help of two independently
written computer programs, which realize the Jacobi polynomial
expansion method.
The first Program has been created as the result
of the works of Ref.\cite{Dubna}, while the second Program previously
found its more distinguished application in the process of the fits
of Ref.\cite{PKK}. The Jacobi polynomial parameters were
fixed as $\alpha=0.7$ and $\beta=3$ like in our previous paper
of Ref.\cite{our4}. Both Programs are giving the identical
results,
which are presented in Table 1. All fits were done in the LO, NLO and
NNLO level both without twist-4 contributions and with twist-4
corrections, taken into account through the expression of Eq.(4).

In Table 2 the NLO results, obtained with
taking into account the IRR-model estimates of
the  twist-4
contributions, are compared to the outcomes of the independent fits
of the CCFR collaboration, which are presented in  the second
reference from Ref. \cite{CCFRrev}. Let us remind that these results
were got using the DGLAP-equation. One can see that within the
statistical uncertainties our NLO results agree with the results
obtained by the CCFR collaboration
with the help of another method and
another computer program.

Looking carefully on Table 1 we arrive to the following conclusions:
\begin{itemize}
\item The effects of the NNLO perturbative QCD contributions turn
out to be very important in the analysis of the CCFR'97 data
for the $xF_3$ structure function. Indeed, for
the different $Q^2$-cuts
they are diminishing the values of the QCD scale parameter
$\Lambda_{\overline{MS}}^{(4)}$ by over $50-130~MeV$ (!)
provided the twist-4 corrections are taken into account through the
IRR-model of Ref.\cite{DW};
\item For the low $Q^2$-cuts $Q^2>1~GeV^2$ and $Q^2>5~GeV^2$
we observe the existence of the following hierarchy
$(\chi^2)_{NNLO}<(\chi^2)_{NLO}<(\chi^2)_{LO}$, which demonstrate
the importance of taking into account of the effects of perturbative
QCD corrections in the process of the
fits of the concrete
experimental data for the structure functions of DIS;
\item When the effects of higher-twist corrections are
included,
the NNLO results for $\Lambda_{\overline{MS}}^{(4)}$ are
less sensitive to the values of $Q^2$-cuts of the CCFR'97
experimental data  than the similar
NLO results;
\item The inclusion of the twist-4 contributions into the fits of
the CCFR'97 data results in the detectable increase of the central
values of $\Lambda_{\overline{MS}}^{(4)}$ at the NLO for almost
all  $Q^2$-cuts  besides the cut
$Q^2>1~GeV^2$ (to our point of view,  the  analysis of the
data in the latest region can be more
sensitive to the change of  the value of the
factorization scale $Q_0^2=5~GeV^2$
say to the value $Q_0^2=1~GeV^2$, which is more appropriate for
fitting the data in the region of momentum transfered $Q^2>1~GeV^2$);
\item At the NLO level the values of the fitted parameter
$A_2'$, responsible for the inclusion of the twist-4 corrections,
do not contradict the value $A_2'\approx -0.2~GeV^2$, which was chosen
in the work of Ref.\cite{DW} for the parametrization of the twist-4
corrections within the IRR model approach;
\item
We found that at the NNLO level the IRR-motivated expression
of the twist-4 contributions influence the outcomes
of the fits less vividly than at the NLO level.
This feature is related to the fact, that the central values
of the parameter $A_2'(HT)$ which result from the NNLO fits
turn out to be rather small.
Moreover, we even are unable to fix the sign
of $A_2'(HT)$ at the NNLO level, since the statistical
error-bars of the value of this parameter are large (!!).
This observed property might reveal the feature of the partial
cancellation of the IRR contribution to the twist-4 correction
of the DIS structure function with the possible part of the
$O(1/Q^2)$-term, which can arise from the summation of the
ultraviolet renormalon effects, discussed recently
e.g. in Ref.\cite{AZ2} and partly taken
into account in our analysis through the certain sets of the
perturbative NNLO corrections.
\end{itemize}

However,
one should keep in mind the results of
the recent quantum mechanical consideration of Ref.\cite{PP},
which are
demonstrating that the renormalons might give a
noncomplete information on the structure of nonperturbative effects.
In its turn this can indicate that the IRR higher-twist contribution
to Eq.(4) might have non-controllable theoretical uncertainties.
These interesting conclusions make very  important the
direct extraction of the information about
the higher-twist contributions from the analysis of the
CCFR'97 experimental data
following the lines of the work of
Ref.\cite{Sasha}, devoted to the similar analysis of
the old CCFR data.

{\bf 3.}~
In this section we  follow the ideas of
Ref.\cite{VM,Sasha,Sasha2} and
undertake the simultaneous extraction
of the parameters $A,b,c,\gamma$,
$\Lambda_{\overline{MS}}^{(4)}$
and the $x$-shape of the twist-4 corrections
from the CCFR'97 $xF_3$
experimental data
using the model of Eq.(6).
In accordance with the
$x$-bin structure of the  CCFR'97 data we
consider 16 bins $x_i$, presented in Table 3.
We have taken
the number of active flavours $f=4$ in all region of momentum
transfered $5~GeV^2\leq Q^2\leq 199.5~GeV^2$. The results of the
LO, NLO and NNLO fits, performed in the kinematical
conditions $Q_0^2=5~GeV^2$ and $Q^2>5~GeV^2$,
are presented in Tables 3,4 and Figs.1-3. At
Fig.2
we also plot the prediction for $h(x)$, obtained
within the framework of the IRR
model approach of Ref.\cite{DW}. Note, that in the work of
Ref.\cite{DW} the concrete predictions were obtained for the twist-4
contribution, written down in the slightly different form:
$xF_3^{HT}(x,Q^2)= xF_3^{LT}D(x)/Q^2$.
To transform this prediction to our normalization of
Eq.(6) we used the program to compute the $x$-shape of this power
term, written down by the authors of
Ref.\cite{DW}\footnote{We are
grateful to B.R. Webber for providing us this program}.

Let us now describe the main conclusions, which follow
from
the results of Tables 3,4.
\begin{itemize}
\item The statistical uncertainties of the values of
$\Lambda_{\overline{MS}}^{(4)}$ are increasing drastically.
This property results from  increasing of the number of the
fitted parameters and reflects the correlation of the uncertainties
of the values of the parameters $A,b,c,\gamma$ and
$\Lambda_{\overline{MS}}^{(4)}$ with the ones, which appear in view
of the lack of the
precise theoretical information about the twist-4
contributions;
\item The non-normalized to the number of degrees of freedoms value of
the $\chi^2$-parameter drastically decreases
and turn out to be rather stable to the inclusion of the
NLO and NNLO corrections into the fits of the CCFR'97 experimental
data. This property is
reflecting the positive feature of the incorporation of the
additional 16 parameters $h(x_i)$ into the analysis of the
CCFR'97 experimental data;
\item We
confirm the conclusion of Sec.2 about the importance of taking into
account NNLO contributions to Eq.(2)
in the concrete fits of the experimental
data. Indeed, the NNLO QCD corrections are decreasing
both the central value of the parameter
$\Lambda_{\overline{MS}}^{(4)}$ and the corresponding statistical
uncertainties. Moreover, making additional fits, we found that the
strong decrease of the value of the parameter
$\Lambda_{\overline{MS}}^{(4)}$ at the NNLO is resulting from the
incorporation of the NNLO corrections into Eq.(2), but not into the
relation between $\alpha_s(Q^2)$ and $\Lambda_{\overline{MS}}^{(4)}$;
\item We have checked that the results of the LO and NLO fits are
almost non-sensitive to the changes of $N_{max}=10$ in Eq.(3) to
$N_{max}=6$, considered in the process of the NNLO fits. This feature
is rather welcomed from the point of view of the confirmation of the
reliability of the application of the Jacobi polynomial expansion
machinery;
\item At the NLO and in the region of the intermediate
values of $x$ the form of the twist-4 correction $h(x)$ turns out to
be in satisfactory agreement with the $x$-behavior of the IRR-model
prediction of Ref.\cite{DW}. This property is confirming the
conclusion of
Ref.\cite{Stein}, that in the first approximation the IRR model
does in fact good job as far as $x$-dependence of the higher-twist
corrections is concerned;
\item However, we also confirm the
conclusion of Refs.\cite{PP,Stein} that the IRR model might give the
non-complete estimate of the mass scale of the higher-twist
contributions. Indeed, we observe that the inclusion of the NNLO
corrections into the fits of the experimental data is resulting in
the less vivid  $x$-dependence  of the twist-4
correction $h(x)$ (!), which model the effects of the truncation of
the perturbative QCD expressions for the related moments
$M_n^{NS}(Q^2)$ and of the reconstructed structure function
$xF_3^{LT}(x,Q^2)$ at the definite order of perturbation theory.
Moreover, looking carefully on Table 3 we observe
that in almost all $x_i$-bins the following tendency
$|h^{cv}(x_i)|_{NNLO}\leq |h^{cv}(x_i)|_{NLO}$ is revealed, where
$h^{cv}(x_i)$ denotes the central values of the corresponding
twist-4 parameters.
These
observations are one of the main outcomes of our analysis, presented
in this Section. Following the hypothesis, made in the item 6 of
Sec.2, we are relating the explanation of this feature to the
possible NNLO manifestation of the summed up ultraviolet renormalon
contributions, discussed in detail in Ref.\cite{AZ2};
\item In the region of the intermediate values of $x$ the results
of our extarctions of $h(x_i)$ (see Figs.1-3) do not contradict
to the predicted in Ref.\cite{Br,Nason} behaviour $|h(x)|\rightarrow
(1-c)^{c-1}$ (where we find $3<c\leq 4$). Note, however, that the
predictions of Ref.\cite{Br,Nason} were obtained in the asymptotic
limit $x_i\rightarrow 1$, while our results are related to the
kinematical region $0.0125\leq x_i\leq 0.65$, which in fact is lying
quite far from this asymptotic regime.
\item The results of the fits demonstrate that
the twist-4 contributions can have the sign-alternating character,
discussed for the first time in the process of the fits of the DIS
experimental data in Ref.\cite{PSI}, where another phenomenological
approach was used for modeling higher-twist contributions.
It should be stressed, that our picture of the $x$-behaviour of
$h(x)$ is more rich.  Indeed, while the considerations of
Ref.\cite{PSI} indicate the existence of only one zero in the
values of $h(x)$, we see that $h(x)$ can have two zeros:
the first one at small values of $x$ and another one in the
region of larger values of $x$ (see Fig.1-Fig.3).
\end{itemize}

{\bf 4.}~In Ref.\cite{CCFRrev} the final result for the parameter
$\Lambda_{\overline{MS}}^{(4)}$, extracted from the $xF_3$
CCFR'97 data, was presented for the case
of the fits, made within the kinematical conditions
$Q_0^2=5~GeV^2$
and $Q^2>5~GeV^2$. It was obtained with the help of the
newly proposed non-conventional procedure of the treatment
of the systematic uncertainties, called as the "global systematic
fit". It is based on the incorporation of the systematic variations
of the structure functions as the part of the fit and on the
redefinition of the $\chi^2$-parameter. The application of this
procedure results in the substantial reduction of the value of the
newly defined $\chi^2$-parameter
and of the combined statistical and systematic error bars. However,
in order to get the
feeling, whether the theoretical error of the NLO value of
$\Lambda_{\overline{MS}}^{(4)}$ was
estimated in the process of the work of Ref.\cite{CCFRrev} correctly,
we will compare the outcomes of our NLO and NNLO studies to the
following CCFR'97 result, obtained in the second work from
Ref.\cite{CCFRrev}:
\begin{equation}
NLO:~~~\Lambda_{\overline{MS}}^{(4)}=387\pm 42(stat)\pm 93(syst)
\pm 17(HT) \pm 70 (scale)~~~MeV
\label{CCFR97}
\end{equation}
where the statistical and systematic uncertainties were determined
by the conventional way and the higher-twist error was estimated by
varying  the IRR model parameter $A_2'$ in the following
interval $(0,-0.2~GeV^2)$ with the central value
$A_2'=A_2'(DW)/2=-0.1~GeV^2$ \footnote{The abbreviation $A_2'(DW)$
corresponds to the exact parametrization of the twist-4 contribution
through the IRR-model of Ref.\cite{DW}}.

Notice, that in Eq.(7) the scale-scheme
uncertainties, which in part reflect the sensitivity
of the outcomes of the fits to the non-taken into account
higher order QCD corrections,
were {\it ad hoc} taken from
the estimate $\Delta\alpha_s(M_Z)=\pm 0.004$, made in the process
of the NLO studies of Ref.\cite{VM}, aimed to the NLO analysis of the
BCDMS $F_2$ structure function data of the $\mu N$ DIS.
Evolving now the result of Eq.(7) through the threshold of the
production of the $b$-quark with the help of
the derived in Ref.\cite{Marciano} relations with the choice
$m_b\approx 4.6~GeV$, we get the following value of
$\alpha_s(M_Z)$:
\begin{equation}
NLO:~~~\alpha_s(M_Z)=0.122\pm 0.002
(stat) \pm 0.005 (syst) \pm 0.001 (HT) \pm 0.004 (scale)~~~.
\label{CCFR97res}
\end{equation}
which corresponds to analysis of the CCFR'97 $xF_3$ experimental
data,  performed in the second work of Ref.\cite{CCFRrev}
with the  separate determination of the statistical and systematic
experimental uncertainties.

Since we are able to incorporate in our fits the effects of the
NNLO corrections and extract the value of $A_2'$-- parameter (see Sec.2),
we can  get the
feeling on the validity of the estimates of the theoretical
uncertainties in Eq.(7) and Eq.(8). Indeed, for $Q_0^2=5~GeV^2$
and $Q^2>5~GeV^2$ our NLO and NNLO results, which follow from the
ones of Table 1, read
\begin{eqnarray}
NLO~~HT~ of~ Ref.\cite{DW}:~~
\Lambda_{\overline{MS}}^{(4)}&=& 371\pm 31(stat)\pm
93(syst) ~MeV\\
\nonumber
NNLO~~HT~ of~ Ref.\cite{DW}:~~
\Lambda_{\overline{MS}}^{(4)}&=&
293\pm 29(stat)\pm 93(syst) ~MeV
\end{eqnarray}
where the
systematic uncertainties are taken from the original analysis
of Ref.\cite{CCFRrev} (which we are unable neither to check nor to
improve) and the statistical uncertainties
in Eq.(9) are taking into account
the deviation of the twist-4 parameter $A_2'$ in
Eq.(4) from its canonical value of Ref.\cite{DW}.

One can see that
the inclusion of the NNLO QCD corrections into the fits of the
CCFR'97 experimental data decreases the central value of the parameter
$\Lambda_{\overline{MS}}^{(4)}$ by over $80~MeV$. This amount is
comparable with the estimate of the scale-scheme dependence
uncertainty in Eq.(7).  In its turn, this fact
means that the theoretical uncertainties
in the NLO results of Eq.(7) and Eq.(8) are underestimated.

Indeed, it was already
demonstrated in the second work from Ref.\cite{CCFRrev}
that the outcomes of the NLO fits of the CCFR'97 experimental
data for the $xF_3$ structure function turn out to be
sensitive to the upper value of the $x$-cut.
The variation of this cut from $x<0.7$ to $x<0.6$
and to $x<0.5$ (which is equivalent to  neglecting the high
$Q^2$-data for the $xF_3$ structure function)
is decreasing the central value of the CCFR'97 NLO result
for $\Lambda_{\overline{MS}}^{(4)}$ by over $32~MeV$ and
$40~MeV$ correspondingly\footnote{Note, that for the cut $x<0.6$ we
confirmed this decreasing tendency.}. Moreover, for the
cut $x<0.7$ there is still the
place for other theoretical uncertainties
(besides the controlled by us effects of the NNLO
corrections), which appear due to the factorization scale
dependence, the unknown $N^3LO$ contributions and the
sensitivity to nuclear effects.
Since
the CCFR'97 data for the $(xF_3)^{Fe}$ structure function were not
corrected by the latter ones, we made the fits, analogous to the
ones described in Sec.3, but with modeling the ratio
$R_{Fe}=F_3^{Fe}/F_3^N$ by the ratio
$R_D=F_3^D/F_3^N$, which was fixed by the approach,
considered in Ref.\cite{ST}. These fits also
result in the effect of the decreasing of the value of the parameter
$\Lambda_{\overline{MS}}^{(4)}$, but by less considerable amount
of over $15~MeV$. It is highly desirable to check this
conclusion by using some other approach for modeling the value
of $R_{Fe}$, say the one, developed in Ref.\cite{Kulagin} in the case
of the analysis of the $F_2$ structure functions data.

Evolving now the results of Eq.(9) through the thresholds of the
production of the $b$-quark and substituting the values of the
parameter $\Lambda_{\overline{MS}}^{(5)}$ into the inverse-log
expression for $\alpha_s(M_Z)$, we get
\begin{eqnarray}
NLO~~HT~ of~ Ref.\cite{DW}:~~
\alpha_s(M_Z)&=& 0.121\pm 0.002(stat)\pm
0.005(syst)\pm 0.006(theory)\\
\nonumber
NNLO~~HT~ of~ Ref.\cite{DW}:~~
\alpha_s(M_Z)&=&
0.117\pm 0.002(stat)\pm 0.005(syst)\pm 0.003 (theory)~~.
\end{eqnarray}
The theoretical error bars of the NLO and NNLO values of
$\alpha_s(M_Z)$ in Eq.(10) are taking into account all discussed
uncertainties plus the arbitrariness of application of the
procedure of passing threshold of the production of $b$-quark,
which following the considerations of Ref.\cite{ShMS} we estimate
as $\Delta\alpha_s(M_Z)\pm 0.001$.

The NNLO results from Eq.(10) turns out to be in agreement with
the following NNLO value of $\alpha_s(M_Z)$:
$\alpha_s(M_Z)=
0.115\pm 0.001(stat)\pm 0.005(syst)\pm 0.003 (twist)\pm
0.0005(scheme)$,
which was extracted in Ref.\cite{CK}
from the comparison of the old CCFR
data for the Gross-Llewellyn Smith sum rule at the reference
point $Q^2=3~GeV^2$ \cite{CCFRGLS} with the results of the
perturbative NLO and NNLO calculations of Ref.\cite{GL},
supplemented with the information about the
concrete values of the twist-4
corrections, obtained in  Ref.\cite{BK}. The optimistic
scheme-dependence error in the NNLO result of Ref.\cite{CK} is
reflecting the uncertainty due to the unknown effects of the higher
$N^3LO$ perturbative QCD corrections, estimated later on in the works
of Ref.\cite{KS}. It should be stressed, however, that in view of the
changes of the CCFR $\nu N$ data the experimental analysis of
Ref.\cite{CCFRGLS} should be updated.

Let us also mention, that the discussed in Sec.3 fits of the
CCFR'97 data, which have the aim to extract the $x$-shape of the
twist-4 contributions and to verify the predictions of the
IRR-model, give as the outcomes the comparable values of
$\Lambda_{\overline{MS}}^{(4)}$ and thus $\alpha_s(M_Z)$. Indeed,
when transformed through the thresholds of the production of the
$b$-quark, the  NLO and NNLO results of Table 4
\begin{eqnarray}
NLO~~HT~free:~~
\Lambda_{\overline{MS}}^{(4)}&=& 428\pm 158(exp)
 ~MeV\\
\nonumber
NNLO~~HT~free:~~
\Lambda_{\overline{MS}}^{(4)}&=&
264\pm 85(exp) ~MeV
\end{eqnarray}
are equivalent to the following values of $\alpha_s(M_Z)$:
\begin{eqnarray}
NLO~~HT~ free:~~
\alpha_s(M_Z)&=& 0.124\pm 0.007(exp)\pm
 0.010(theory)\\
\nonumber
NNLO~~HT~free:~~
\alpha_s(M_Z)&=&
0.115\pm 0.006(exp)\pm 0.003 (theory)~~,
\end{eqnarray}
where the experimental uncertainties are estimated by the statistical
errors (which are increased due to the inclusion of the additional 16
twist-4 parameters $h(x_i)$ into the fits), the NLO  theoretical
errors are fixed by the difference between the NLO and NNLO
results and the remaining NNLO theoretical ambiguities are determined
by the uncertainties, taken by us into account in the process of
estimates of the theoretical ambuguities of the NNLO value of
$\alpha_s(M_Z)$, presented in Eq.(10).

It should be stressed, that
our results of Eqs.(10),(12) do not contradict to the NLO value
$\alpha_s(M_Z)=0.113\pm 0.003(exp)\pm 0.004(theory)$,
obtained in the process of the NLO analysis of the BCDMS and SLAC
data for the $F_2$ structure function of $\mu N$ and $eN$ DIS
\cite{VM}.
However, there are still the number of questions to be answered in
future. The first, theoretical one, is related to the necessity of
the inclusion into the analysis of the BCDMS and SLAC data the
NNLO perturbative QCD corrections to the singlet part of the
$F_2$ structure function (we consider the results of the NNLO
non-singlet fits of Ref.\cite{PKK} still non-complete, since in the
process of these fits only the data for $x>0.35$ was analyzed
without taking into account target mass corrections).
The second, pure experimental, question is related to the problem
whether the scientific community will accept the proposed in
Ref.\cite{CCFRrev} new approach to treating experimental
uncertainties. Note, that the accepted at present conventional
way of their definitions is leading to the conclusion, that
the combination of the traditionally defined statistical and
systematic uncertainties of the CCFR'97 experimental data is not
too small. Keeping this in mind, we agree with the conclusions of
the analysis of the experimental data for the Bjorken polarized sum rule
\cite{ABFR,Ioffe} that in order to extract the more precise value
of $\alpha_s(M_Z)$ from the characteristics of DIS with taking into
account both the higher-order perturbative QCD corrections and
higher-twist effects it is necessary to have more precise
experimental data. This conclusion is supporting the detailed
investigation of the third problem, revealed in the process of
the analysis of the CCFR'97 experimental data of Ref.\cite{CCFRrev}.
It is related to the necessity of more detailed
investigation of the possible discrepancies between the CCFR'97
$F_2$ $\nu N$ data in the region of small $x$ ($x\approx
0.01-0.07$) and the similar experimental data for the
$F_2$ structure functions of the DIS of charged leptons on
nucleons, measured by the BCDMS, NMC and SLAC collaborations.
Among the possible ways of attacking this
experimental problem are the possible more detailed studies of $\nu
Al$ DIS data of the IHEP-JINR neutrino detector (for the already
available preliminary results see Ref.\cite{IJ}) and the
continuation of the careful measurements of the structure functions
of $\nu Fe$ DIS at Fermilab Tevatron, which is the main goal of the
NuTeV collaboration.

{\bf Acknowledgments}

We are grateful to  M.H.~Shaevitz and W.G.~Seligman for
providing us the available CCFR data of Ref.\cite{CCFRrev}
and to W.G.~Seligman for the detailed discussions of the
outcomes of the DGLAP fits of the CCFR'97 data.

We wish  to thank
G.~Altarelli, J.C.~Collins, A.V.~Efremov, E.A.~Kuraev, L.N.~Lipatov,
P.~Nason, A.A.~Pivovarov,
I.A.~Savin, M.A.~Shifman,
B.R.~Webber and F.J.~Yndurain for useful discussions.

This work is supported by the Russian Fund for Fundamental Research,
Grant N 96-02-18897. The work of two of us (A.L.K. and A.V.S.)
was done within the scientific program of the INTAS project N
93-1180. The work of G.P. was supported by CICYT
(Grant N AEN96-1773) and
Xunta de Galicia (Galician Autonomous Government, Spain)
Grant N XUGA-20604A96.

\newpage
\begin{tabular}{||c||c|c|c|c|c|c|c||} \hline
\hline
                $ Q^2 > $           &
                $\Lambda_{\overline{MS}}^{(4)}$ (MeV)     &
                     A              &
                     b        &
                     c         &
                $ \gamma $          &
                $ A_2^\prime$(HT)   &
                $ \chi^2$/nep
                   \\
\hline\hline
 1 & & & & & & &  \\
LO & 277$\pm$31 & 4.65$\pm$0.14 & 0.69$\pm$0.01 & 3.90$\pm$0.04 &
  1.82$\pm$0.13 & -- & 146.0/110  \\

   & 367$\pm$43 & 3.95$\pm$0.40  & 0.63$\pm$0.03 & 3.87$\pm$0.05 &
  2.48$\pm$0.47 & -0.25$\pm$0.04 & 114.4/110  \\

NLO & 322$\pm$27 & 3.67$\pm$0.72 & 0.62$\pm$0.05 & 3.76$\pm$0.10 &
  2.32$\pm$0.94 & -- & 114.1/110  \\

   & 322$\pm$23 & 3.61$\pm$0.11  & 0.61$\pm$0.01 & 3.76$\pm$0.04 &
  2.39$\pm$0.12 & -0.07$\pm$0.04 & 111.4/110  \\

NNLO & 275$\pm$24 & 4.01$\pm$0.34 & 0.64$\pm$0.02 & 3.61$\pm$0.06 &
  1.56$\pm$0.33 & -- & 104.9/110 \\

   & 276$\pm$25 & 4.00$\pm$0.34  & 0.64$\pm$0.02 & 3.61$\pm$0.06 &
  1.56$\pm$0.32 & 0.01$\pm$0.04 & 104.9/110  \\
\hline
 5 & & & & & & &  \\
LO & 266$\pm$37 & 5.13$\pm$0.46 & 0.72$\pm$0.03 & 3.87$\pm$0.05 &
  1.42$\pm$0.33 & -- & 113.2/86  \\

   & 436$\pm$56 & 4.73$\pm$0.39  & 0.68$\pm$0.03 & 3.79$\pm$0.05 &
  1.82$\pm$0.31 & -0.33$\pm$0.06 & 82.8/86  \\

NLO & 341$\pm$41 & 4.05$\pm$0.38 & 0.65$\pm$0.03 & 3.71$\pm$0.06 &
  1.96$\pm$0.36 & -- & 87.1/86  \\

   & 371$\pm$31 & 3.94$\pm$0.12  & 0.64$\pm$0.01 & 3.69$\pm$0.04 &
  2.05$\pm$0.12 & -0.12$\pm$0.05 & 82.4/86  \\

NNLO & 293$\pm$29 & 4.25$\pm$0.38 & 0.66$\pm$0.03 & 3.56$\pm$0.07 &
  1.33$\pm$0.33 & -- & 78.4/86  \\

   & 293$\pm$29 & 4.25$\pm$0.37  & 0.66$\pm$0.03 & 3.56$\pm$0.07 &
  1.33$\pm$0.32 & -0.01$\pm$0.05 & 78.4/86  \\
\hline
 10 & & & & & & &  \\
LO & 287$\pm$37 & 6.52$\pm$0.20 & 0.79$\pm$0.01 & 3.73$\pm$0.04 &
  0.60$\pm$0.10 & -- & 77.7/63  \\

   & 531$\pm$74 & 6.48$\pm$0.65  & 0.79$\pm$0.04 & 3.68$\pm$0.05 &
  0.94$\pm$0.26 & -0.52$\pm$0.11 & 58.0/63  \\

NLO & 350$\pm$36 & 5.29$\pm$0.21 & 0.74$\pm$0.01 & 3.64$\pm$0.06 &
  1.06$\pm$0.17 & -- & 64.4/63  \\

   & 439$\pm$55 & 4.92$\pm$1.06  & 0.73$\pm$0.07 & 3.61$\pm$0.08 &
  1.42$\pm$0.75 & -0.24$\pm$0.10 & 58.6/63  \\

NNLO & 308$\pm$34 & 4.65$\pm$0.53 & 0.70$\pm$0.04 & 3.52$\pm$0.08 &
  1.06$\pm$0.38 & -- & 58.7/63  \\

  & 313$\pm$37 & 4.64$\pm$0.52 & 0.70$\pm$0.04 & 3.52$\pm$0.08 &
  1.07$\pm$0.38 & -0.03$\pm$0.09 & 58.6/63  \\
\hline
 15 & & & & & & &  \\
LO & 319$\pm$41 & 7.58$\pm$0.24 & 0.84$\pm$0.01 & 3.61$\pm$0.05 &
  0.16$\pm$0.09 & -- & 58.8/50 \\

   & 531$\pm$57 & 7.76$\pm$0.26  & 0.86$\pm$0.02 & 3.61$\pm$0.04 &
  0.45$\pm$0.09 & -0.57$\pm$0.13 & 50.2/50  \\

NLO & 365$\pm$40 & 6.28$\pm$0.21 & 0.80$\pm$0.02 & 3.60$\pm$0.04 &
  0.58$\pm$0.09 & -- & 53.0/50  \\

   & 441$\pm$44 & 5.98$\pm$0.19  & 0.80$\pm$0.02 & 3.57$\pm$0.05 &
  0.87$\pm$0.10 & -0.25$\pm$0.13 & 50.9/50  \\

NNLO & 314$\pm$37 & 4.68$\pm$0.75 & 0.70$\pm$0.05 & 3.53$\pm$0.08 &
  1.08$\pm$0.51 & -- & 50.9/50 \\

   & 308$\pm$45 & 4.69$\pm$0.72  & 0.70$\pm$0.05 & 3.53$\pm$0.08 &
  1.07$\pm$0.50 &  0.03$\pm$0.14 & 50.8/50  \\
\hline \hline
\end{tabular}
{{\bf Table 1.} The results of the fits of the $xF_3$ ($\nu$Fe)
revised data
of the  CCFR collaboration. The values of $Q^2$ and $A_2^{'}(HT)$ are given in
GeV$^2$. The statistical errors are presented.
($\chi^2$ is normalized  to the
number of experimental points (nep).)}

\begin{center}
\begin{tabular}{|c|c|c||c|c||} \hline
 $Q^2$ cut (GeV$^2$)& $x$ cut & Order &
 $\Lambda_{\overline{MS}}^{(4)}$ (MeV)
& $\chi^2$/nep  \\
\hline
$Q^2>5 $ & $x<0.7$     & NLO & $371 \pm 30$ & $82.4/86$ \\
& CCFR result &     & $387 \pm 42$ & $81.8/86$\\
  $Q^2>10$ & $x<0.7$     & NLO & $439 \pm 55$ & $58.6/63$\\
& CCFR result &     & $422 \pm 47$ & $58.0/63$ \\
$Q^2>15$ & $x<0.7$
& NLO & $441 \pm 44$  &$50.9/50$ \\ & CCFR result &     &
$433 \pm 55$ & $50.0/50$ \\
\hline
\end{tabular}
\end{center}
{{\bf Table 2.} The
comparison of the results of our NLO fits with the
revised results provided
by the CCFR collaboration in Ref.[3]}

\newpage

\begin{center}
\begin{tabular}{||c|c|c|c||} \hline \hline
                &      LO              &            NLO       &   NNLO \\ \hline
 $\chi^2$/nep    &  66.2/86           &        65.6/86  &  65.7/86
 \\ A &   5.44 $\pm$   1.74   &   3.70 $\pm$  1.56    &    4.54
  $\pm$   0.88   \\ b &  0.74 $\pm$  0.10  &  0.66  $\pm$  0.11    &
  0.69 $\pm$    0.06  \\ c &   4.00 $\pm$   0.18   &   3.78  $\pm$
  0.21    &   3.72 $\pm$    0.19   \\ $\gamma$  &1.72 $\pm$  1.25  &
2.86 $\pm$  1.72   &   1.43  $\pm$    0.69 \\
\hline \hline
   $x_i$                     &\multicolumn{3}{c||}{  $h(x_i)~[GeV^2]$ }                                  \\  \hline
0.0125 &   0.206 $\pm$ 0.321 &    0.213  $\pm$ 0.332 &     0.170 $\pm$ 0.302         \\
0.0175 &   0.061 $\pm$ 0.268 &    0.091  $\pm$ 0.289 &     0.030 $\pm$ 0.241       \\
0.025  &   0.146 $\pm$ 0.204 &    0.220  $\pm$ 0.241 &     0.136 $\pm$ 0.178         \\
0.035  &  -0.021 $\pm$ 0.185 &    0.114  $\pm$ 0.240 &     0.008 $\pm$ 0.179       \\
0.050  &   0.031 $\pm$ 0.142 &    0.245  $\pm$ 0.230 &     0.124 $\pm$ 0.164       \\
0.070  &  -0.145 $\pm$ 0.127 &    0.138  $\pm$ 0.233 &     0.033 $\pm$ 0.165       \\
0.090  &  -0.177 $\pm$ 0.125 &    0.139  $\pm$ 0.225 &     0.076 $\pm$ 0.166       \\
0.110  &  -0.340 $\pm$ 0.126 &   -0.015  $\pm$ 0.205 &    -0.026 $\pm$ 0.166        \\
0.140  &  -0.404 $\pm$ 0.114 &   -0.092  $\pm$ 0.147 &    -0.027 $\pm$ 0.141       \\
0.180  &  -0.350 $\pm$ 0.164 &   -0.077  $\pm$ 0.122 &     0.054 $\pm$ 0.127       \\
0.225  &  -0.554 $\pm$ 0.237 &   -0.348  $\pm$ 0.170 &    -0.167 $\pm$ 0.135       \\
0.275  &  -0.563 $\pm$ 0.334 &   -0.462  $\pm$ 0.272 &    -0.196 $\pm$ 0.193       \\
0.350  &  -0.314 $\pm$ 0.418 &   -0.368  $\pm$ 0.371 &     0.070 $\pm$ 0.200       \\
0.450  &  -0.117 $\pm$ 0.415 &   -0.266  $\pm$ 0.401 &     0.183 $\pm$ 0.213       \\
0.550  &   0.087 $\pm$ 0.333 &   -0.109  $\pm$ 0.352 &     0.097 $\pm$ 0.234       \\
0.650  &   0.377 $\pm$ 0.215 &    0.221  $\pm$ 0.244 &     0.259 $\pm$ 0.188        \\   \hline  \hline
\end{tabular}
\end{center}
{{\bf Table 3.} The results of the LO, NLO and NNLO QCD fits
of the CCFR'97 $xF_3$ structure functions data
for the values of twist-4 contributions $h(x)$ and for the
parameters $A,b,c,\gamma$ with the corresponding
statistical errors.}
\begin{center}
\begin{tabular}{||c|c|c|c|c|c|c||} \hline \hline
 & \multicolumn{2}{|c|}{$h(x)\neq 0$ free} &
 \multicolumn{2}{|c|}{$h(x)\neq 0$ of Ref.[22]} &
 \multicolumn{2}{|c||}{$h(x)=0$}    \\  \hline
 &  $\chi^2/{nep}$ & $\Lambda_{\overline{MS}}^{(4)}$ $(MeV)$ &
 $\chi^2/{nep}$     & $\Lambda_{\overline{MS}}^{(4)}$ $(MeV)$  &
 $\chi^2/{nep}$ & $\Lambda_{\overline{MS}}^{(4)}$ $(MeV)$  \\ \hline
     LO & 66.2/86  &  338   $\pm$ 169    &  82.8/86 & 436   $\pm$
    56 & 113.3/86  &  266   $\pm$  37 \\
    NLO           &
      65.6/86  & 428    $\pm$ 158    &  82.4/86 & 371   $\pm$  31 &
       87.1/86 &  341   $\pm$  41 \\
       NNLO &    65.7/86  & 264
       $\pm$  85 &  78.4/86         & 293 $\pm$  29   & 78.4/86   &
       293 $\pm$  29 \\ \hline \hline
\end{tabular}
\end{center}
{{\bf Table 4.}
The comparison of the results of the LO, NLO and NNLO fits.
of the CCFR'97 $xF_3$ structure functions data, performed with
different assumptions of the values of the twist-4 parameter $h(x)$.}
\newpage
{\bf Figure captions.}

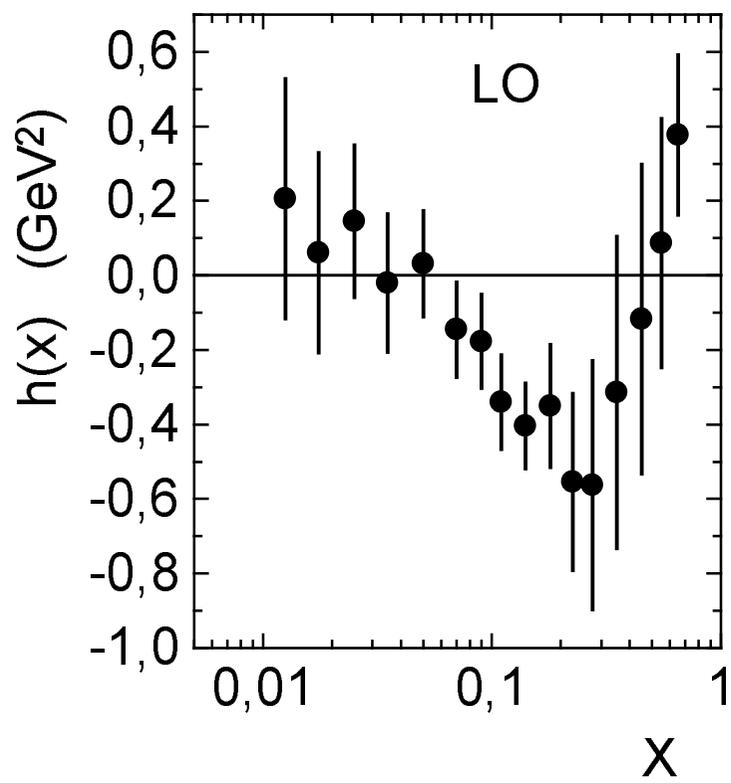
{\bf Fig.1} The results of the LO extraction of the $x$-shape of
the twist-4 contribution $h(x)$.

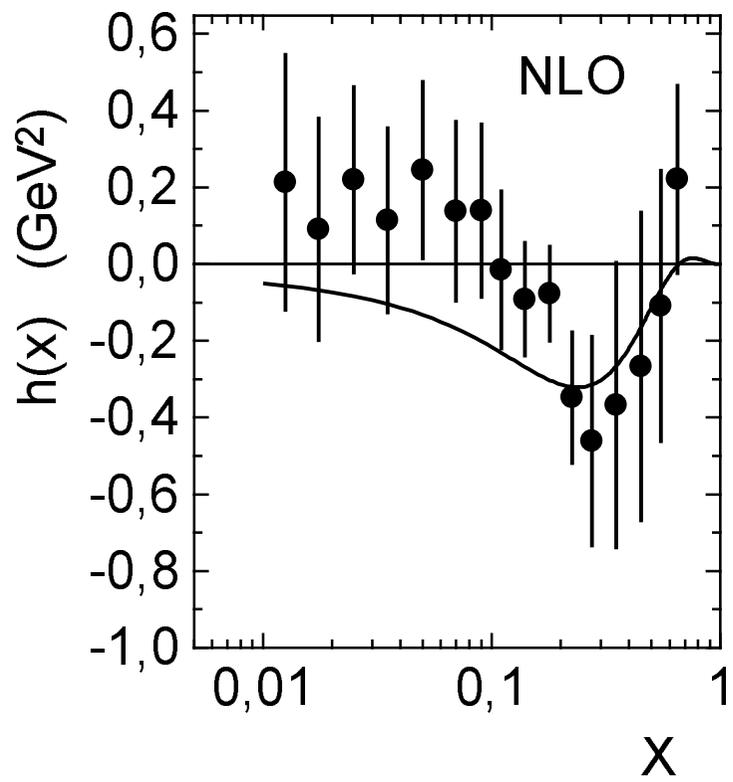
{\bf Fig.2} The results of the NLO extraction of the $x$-shape of
the twist-4 contribution $h(x)$. For the comparison, the IRR-model
prediction of Ref.[22], obtained using  the NLO MRS parametrization,
is also depicted.

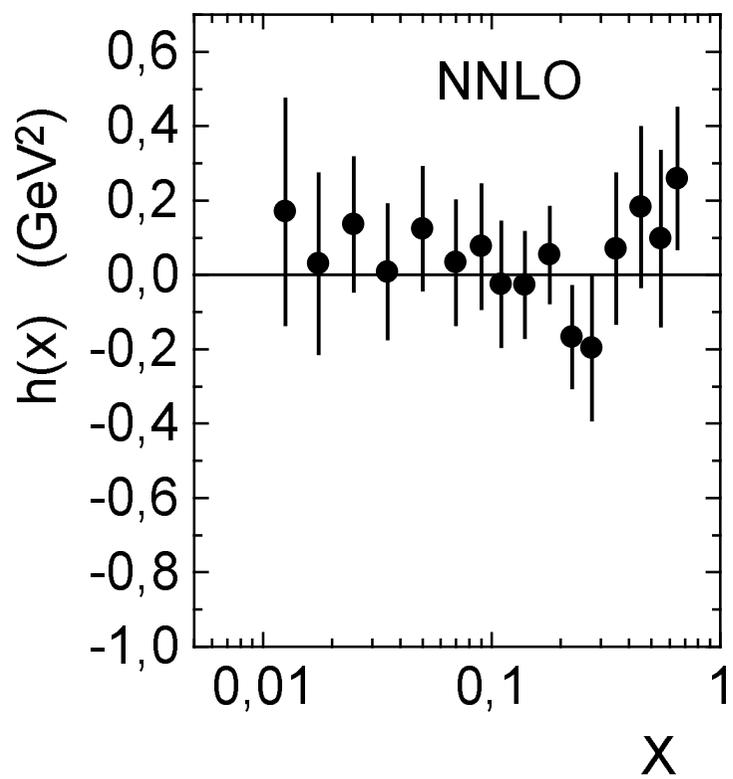
{\bf Fig.3} The results of the NNLO extraction of the $x$-shape of
the twist-4 contribution $h(x)$.

\end{document}